# Toy model for plasmonic metamaterial resonances coupled to two-level system gain

**Martin Wegener\* and Juan Luis García-Pomar[†]**

*Institut für Angewandte Physik and DFG-Center for Functional Nanostructures (CFN), Universität Karlsruhe (TH), Wolfgang Gaede-Strasse 1, D-76128 Karlsruhe, Germany*
[†] *On leave from Instituto de Cienca de Materiales de Madrid, CSIC, Cantoblanco. Madrid E-28049, Spain*
\* *Corresponding Author: Martin.Wegener@physik.uni-karlsruhe.de*

**Costas M. Soukoulis**

*Ames Laboratory and Department of Physics and Astronomy, Iowa State University, Ames, Iowa 50011, U.S.A.*
*and*
*Research Center of Crete, and Department of Materials Science and Technology, 71110 Heraklion, Crete, Greece*

**Nina Meinzer, Matthias Ruther, and Stefan Linden**

*Institut für Nanotechnologie, Forschungszentrum Karlsruhe in der Helmholtz-Gemeinschaft, Postfach 3640, D-76021 Karlsruhe, Germany*

**Abstract:** We propose, solve, and discuss a simple model for a metamaterial incorporating optical gain: A single bosonic resonance is coupled to a fermionic (inverted) two-level-system resonance via local-field interactions. For given steady-state inversion, this model can be solved analytically, revealing a rich variety of (Fano) absorption/gain lineshapes. We also give an analytic expression for the fixed inversion resulting from gain pinning under steady-state conditions. Furthermore, the dynamic response of the "lasing SPASER", i.e., its relaxation oscillations, can be obtained by simple numerical calculations within the same model. As a result, this toy model can be viewed as the near-field-optical counterpart of the usual LASER rate equations.

©2008 Optical Society of America

---

**1. Introduction**

Reducing or compensating the large intrinsic loss of metal-based metamaterials operating at optical frequencies is one of *the* major challenges in the emerging field of photonic metamaterials [1-3]. If this potential show-stopper could be eliminated, many of the envisioned applications such as perfect lenses [4] or cloaking [5,6] at optical frequencies might actually come into reach. For example, the best fabricated negative-index metamaterial structures operating at around 1.4-µm wavelength [7] have shown a figure of merit of FOM=3 (modulus of real to imaginary part of the refractive index). This experimental result can be translated into an absolute absorption coefficient of $\alpha = 3\times10^4$ cm$^{-1}$ = 3 µm$^{-1}$ – which is even larger than the band-to-band absorption of typical direct-gap semiconductors such as, e.g., GaAs (there, $\alpha \approx 10^4$ cm$^{-1}$). At first sight, this level of absorption looks quite depressing as room-temperature steady-state gain coefficients $g=-\alpha$ of this magnitude are not easily achieved at all.

However, an interesting recent theoretical publication by Zheludev et al. [8] – based on the concept of the SPASER (surface plasmon amplification by stimulated emission of radiation) introduced by Stockman et al. [9,10] in 2003 – has essentially shown that it is not the bulk gain coefficient that matters but rather the *effective* gain coefficient of the combined system. Due to pronounced *local-field* enhancement effects in the spatial vicinity of the periodic metallic nanostructure, the effective gain coefficient can be substantially larger than its bulk counterpart. As an extended planar periodic arrangement of identical plasmonic structures that oscillate in phase due to mutual coupling and that exhibit a sub-wavelength period clearly leads to coherent plane-wave emission of light normal to the plane (in close analogy to phased antenna arrays), this two-dimensional active metamaterial structure has been named the "lasing SPASER" [8]. We briefly mention that further theoretical investigations on particular SPASER [11,12] structures, on metamaterials including gain materials [13,14], and on lasing SPASER structures [15] have recently been published. Other work regarding bringing gain to surface plasmons includes, e.g., Refs.[16-19]. Furthermore, recently published experiments on active metal-dielectric nanocavities [20,21] already come fairly close to the original SPASER [9] idea.

In the microscopic calculations by Zheludev et al. [8] based on numerical solutions of the three-dimensional vector Maxwell equations for specific (two-slit) split-ring-resonator nanostructures, the gain medium has been described by a constant (i.e., frequency-independent and emission-independent) negative imaginary part of the gain medium dielectric function. Clearly, it is interesting to investigate modifications of this result

- due to a more realistic frequency-dependent gain,
- due to gain pinning, and
- due to dynamic effects.

Complete microscopic self-consistent numerical calculations of this sort have not been published so far to the best of our knowledge. They may well be feasible, but are certainly rather demanding and will have to deal with the specifics of a particular design.

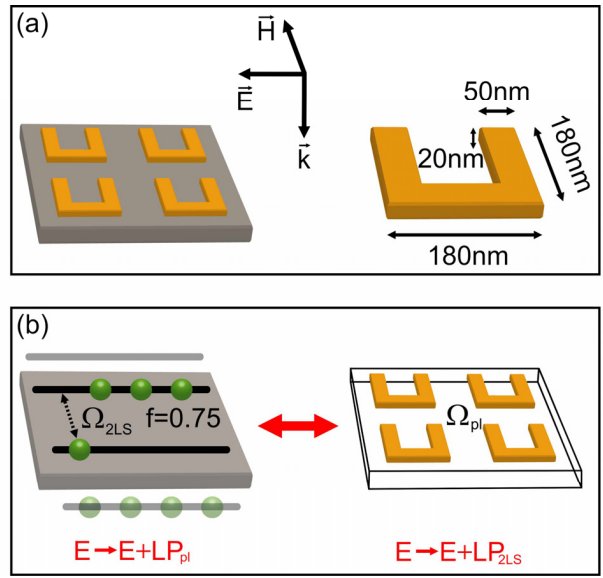

Fig. 1. (a) Illustration of a possible geometry for bringing the optical gain from a thin semiconductor film (bulk, quantum wells, or quantum dots) close to an array of, e.g., plasmonic split-ring resonators. This geometry aims at taking advantage of *local-field*-enhancement effects. The electric-field vector of the incident light lies in the layer plane. (b) Schematic illustration of the toy model for a plasmonic (bosonic) metamaterial resonance coupled to a (fermionic) two-level-system gain resonance via *local-field* interactions. This interaction is described by the phenomenological (Lorentz) parameter $L$.

In this Letter, we rather propose a simple and general toy model based on a fermionic two-level-system resonance (representing the gain medium, e.g., an ensemble of densely spaced semiconductor quantum dots) and a single bosonic resonance (representing the plasmonic resonance of the metamaterial) – connected to each other via a *local-field coupling* analogous to the local-field Lorentz factor. Interestingly, our toy model can be solved analytically for (quasi-) steady-state conditions. Numerical solutions for the time-dependent case are also discussed.

## 2. Definition of the Model

The physics of our toy model is graphically illustrated in Fig.1. The key to the model are the local field (or evanescent-field) interactions, i.e., the local electric field of the two-level system (2LS) is given by the external electric field $E$

$$E(t) = \tilde{E}_0 \cos(\omega t) = \frac{\tilde{E}_0}{2}(\exp(-i\omega t) + \text{c.c.}) \qquad (1)$$

of the light plus a phenomenological constant, $L$, times the polarization of the bosonic mode, $P_{pl}$, i.e.,

$$E \rightarrow E + LP_{pl}. \qquad (2)$$

Correspondingly, the local electric field of the bosonic mode is given by the external electric field plus constant $L$ times the polarization of the 2LS, $P_{2LS}$, i.e.,

$$E \rightarrow E + LP_{2LS}. \qquad (3)$$

As the spatial separation between 2LS and plasmonic resonance is increased (see Fig.1 (a)), the coupling via the plasmonic evanescent field decreases and $L$ is expected to approach zero. Note that we have omitted the self-interactions, which can generally also occur via the local fields. In linear optics, however, self-interaction merely renormalizes the effective resonance center frequency.

Mathematically, the equations of motion for a fermionic two-level system are the famous and well known optical Bloch equations [22]. They can be arranged into the form

$$\dot{p}_{2LS} + (i\Omega_{2LS} + \gamma_{2LS})p_{2LS} = i\hbar^{-1}d_{2LS}E(1-2f), \qquad (4)$$

$$\dot{f} + \Gamma_{2LS}f = i\hbar^{-1}(p^*_{2LS}d_{2LS}E - p_{2LS}d^*_{2LS}E^*). \qquad (5)$$

Here $p_{2LS}$ denotes the (dimensionless) complex transition amplitude, $f = f_{2LS}$ the occupation probability of the upper level, $d_{2LS}$ the dipole matrix element, $E$ the total electric field, $\Omega_{2LS}$ the transition frequency, $\gamma_{2LS}$ the damping (or transverse relaxation) rate, and $\Gamma_{2LS}$ the population (or longitudinal) relaxation rate. The individual electric dipole moment is given by the product $d_{2LS}p_{2LS}$. Multiplying with the volume density of two-level systems, $N_{2LS}$, leads to the macroscopic 2LS polarization $P_{2LS}$, i.e., to $P_{2LS}=N_{2LS}d_{2LS}p_{2LS}+$c.c. The equations of motion for a corresponding single bosonic mode are strictly identical to (4) and (5), except that the factor $(1-2f)$ in equation (4) has to be replaced by unity, i.e.,

$$\dot{p}_{pl} + (i\Omega_{pl} + \gamma_{pl})p_{pl} = i\hbar^{-1}d_{pl}E. \qquad (6)$$

Hence, the equation for the occupation $f_{pl}$ of the bosonic mode becomes irrelevant. In analogy to the 2LS, the macroscopic plasmonic polarization is given by $P_{pl}=N_{pl}d_{pl}p_{pl}+$c.c.

As discussed above, we assume that the coupling between the two resonances is governed by *local-field effects*. Under these conditions, we simply have to replace the electric fields $E$ in (4) and (5) by the term given in (2) and the electric field $E$ in (6) by (3).

We envision that the 2LS is pumped via additional energy levels of the system that are not explicitly accounted for in the 2LS model, resulting in an additional effective pump rate, $\Gamma_{\text{pump}}$, of the 2LS on the right-hand side of (5).

To avoid confusion: Generally, plasmonic nanostructures can have both an electric-dipole and a magnetic-dipole response [1-3]. Here, we have focused on the electric-dipole response (leading to the polarization $P$) as only that can couple directly to readily available gain media at optical frequencies. Thus, our model is not applicable directly to magnetic-dipole plasmonic resonances or to negative-index metamaterials. Yet, it may provide qualitative trends. Our model is applicable directly to, e.g., the electric-dipole (Mie) resonances of the "V" structures envisioned in Ref.[9], but not to the "dark" modes there.

Next, we discuss analytical and numerical solutions of the model defined so far.

## 3. Linear Optical Response for Fixed Occupation

First, we consider the case of a given (i.e., fixed) occupation $f$ of the 2LS and calculate the linear optical response with respect to some probe light with electric field $E$. As long as the system does not exhibit effective stimulated emission, this assumption of fixed $f$ is completely unproblematic. This situation can occur under steady-state pumping conditions indeed. In contrast, as soon as $f$ is such that the system does reveal effective stimulated emission, the situation becomes unstable, hence, it cannot occur under steady-state conditions (also see sections 4 and 5). Still, such values of $f$ can be meaningful under transient conditions, for example, for a probe pulse with electric field $E$ that follows, with some delay, the excitation by a pump pulse at larger photon energies (via the pump rate $\Gamma_{\text{pump}}$ described above).

Under these conditions of fixed $f$, it is straightforward to solve the above coupled equations. Using the usual rotating-wave approximation (RWA) [17], we derive the transition amplitudes $p = \tilde{p}\exp(-i\omega t)$

$$\tilde{p}_{2LS} = \frac{(1-2f)\left(\hbar^{-1}d_{2LS}\dfrac{\tilde{E}_0}{2} + \dfrac{V_{2LS}\hbar^{-1}d_{pl}\dfrac{\tilde{E}_0}{2}}{(\Omega_{pl}-\omega)-i\gamma_{pl}}\right)}{(\Omega_{2LS}-\omega)-i\gamma_{2LS}-(1-2f)\dfrac{V_{pl}V_{2LS}}{(\Omega_{pl}-\omega)-i\gamma_{pl}}}, \tag{7}$$

and

$$\tilde{p}_{pl} = \frac{\hbar^{-1}d_{pl}\dfrac{\tilde{E}_0}{2}+V_{pl}\tilde{p}_{2LS}}{(\Omega_{pl}-\omega)-i\gamma_{pl}}. \tag{8}$$

Here, we have introduced the two resulting coupling frequencies

$$\begin{aligned} V_{2LS} &= \hbar^{-1}d_{2LS}LN_{pl}d_{pl}, \\ V_{pl} &= \hbar^{-1}d_{pl}LN_{2LS}d_{2LS}, \end{aligned} \tag{9}$$

describing the effective back-action of the plasmonic mode onto the 2LS and vice versa, respectively. Note that the numerator of the 2LS transition amplitude in (7) shows the anticipated resonant enhancement of its "oscillator strength", hence of its absorption/gain, via the plasmonic resonance. Strictly speaking, this resonance indexed by "2LS" is a mixed mode composed of the 2LS and the plasmonic resonance. Only for zero coupling, this resonance becomes that of the 2LS. The denominator of (7) also contains a resonance. This aspect can be interpreted as an eigenfrequency and a damping that are effectively frequency dependent, hence, this aspect gives rise to non-trivial lineshapes. As $p_{2LS}$ according to (7) also enters in the numerator of (8), the spectral dependence of $p_{pl}$ is quite complex as well. From this point on, one has two different options to arrive at optical spectra: via a transfer-matrix approach or via a Maxwell-Garnett effective-medium approach.

### 3.1. Transfer-Matrix Approach

The system shown in Fig.1 can be treated as composed of two films with thicknesses $l_{2LS}$ and $l_{pl}$ for the 2LS gain and the plasmonic layer, respectively. The two corresponding dielectric functions and refractive indices result immediately from the two macroscopic polarizations $P_{2LS}$ and $P_{pl}$ given above. Normal-incidence intensity transmittance $T$ and reflectance $R$ spectra can be computed from the well known transfer-matrix approach [26] for layered films. In order to test the relevance of our toy model and in order to adjust the toy model parameters, especially the local-field parameter $L$, we compare such calculations with microscopic numerical calculations on the basis of the complete three-dimensional vector Maxwell equations. These calculations based on the commercially available finite-difference time-domain software package *lumerical* [27] are rather similar to those in Ref.[8], however, with a frequency-dependent two-level-system gain resonance (same parameters as for the toy model) and for plasmonic split-ring resonators (SRR) roughly similar to the ones discussed in Ref.[24]. The SRR are $l_{pl}$=20-nm thick and are arranged on a quadratic lattice with lattice constant $a$=300 nm. The SRR lateral dimensions can be seen from Fig.1 (a). The silver, the SRR are assumed to be composed of, is described by the Drude free-electron model with standard parameters, i.e., plasma frequency $\omega_{pl}$=1.32×10$^{16}$ s$^{-1}$ and collision frequency $\omega_{col}$=1.2×10$^{14}$ s$^{-1}$.

Selected results are depicted in Fig.2, where spectra are arranged as a matrix. Obviously, the agreement between numerical results in column (a) and analytical calculations with fitted parameters in column (b) is very good. In particular, for $f$=0, a pronounced avoided crossing (or "hybridization") of the two resonances occurs, the splitting of which agrees well. This aspect is of utmost importance because the angular frequency splitting is directly related to the quantity $2(V_{2LS}V_{pl})^{1/2}$ – which is proportional to the local-field coupling $L$ (see (9) and (11)). We will see below that the coupling strength is crucial for the lasing SPASER. Furthermore, the overall agreement for $f$=1 is very good as well. Here, the reflectance at its peak largely exceeds unity, indicative of effective gain of the combined system, whereas the transmittance stays well below unity – an aspect that we will explain intuitively below. We can conclude that the local-field approximation is well justified under these conditions.

### 3.2. Maxwell-Garnett Effective-Medium Approach

Alternatively, one can approximate the system shown in Fig.1 as a *single* effective film with thickness $l=l_{2LS}+l_{pl}$. The advantage as compared to the two-layer approach of the previous section is that a single set of optical parameters follows the spirit of a metamaterial and that it allows for obtaining a more intuitive understanding. The corresponding effective parameters can immediately be derived from Maxwell-Garnett effective-medium theory (recall that the electric-field vector is parallel to the plane of the layers, see Fig.1 (a)). In this case, the total effective macroscopic polarization $P$ results as

$$P = \frac{l_{2LS}}{l_{2LS}+l_{pl}} P_{2LS} + \frac{l_{pl}}{l_{2LS}+l_{pl}} P_{pl} = \varepsilon_0 \chi \frac{\widetilde{E}_0}{2} \exp(-i\omega t) + \text{c.c.} \qquad (10)$$

Equation (10) allows for calculating all relevant linear optical properties via the effective optical susceptibility $\chi$, the effective electric permittivity $\varepsilon=1+\chi$, and the effective complex refractive index $n=\pm\varepsilon^{1/2}$. The sign of the root has to be chosen such that Re($n$)≥0. As $\mu=1$ here, we can equivalently say that the sign of the root has to be chosen such that the sign of Im($n$) is the same as the sign of Im($\chi$). The formulas for the normal-incidence intensity transmittance $T$ and reflectance $R$ of a homogeneous slab with complex refractive index $n$ (hence, as $\mu=1$, complex impedance $Z=Z_0/n$, with the vacuum impedance $Z_0$) and thickness $l=l_{2LS}+l_{pl}$ can, e.g., easily be obtained from equations (5.119) and (5.120) of the review article [3] or from optics textbooks (e.g., [26]).

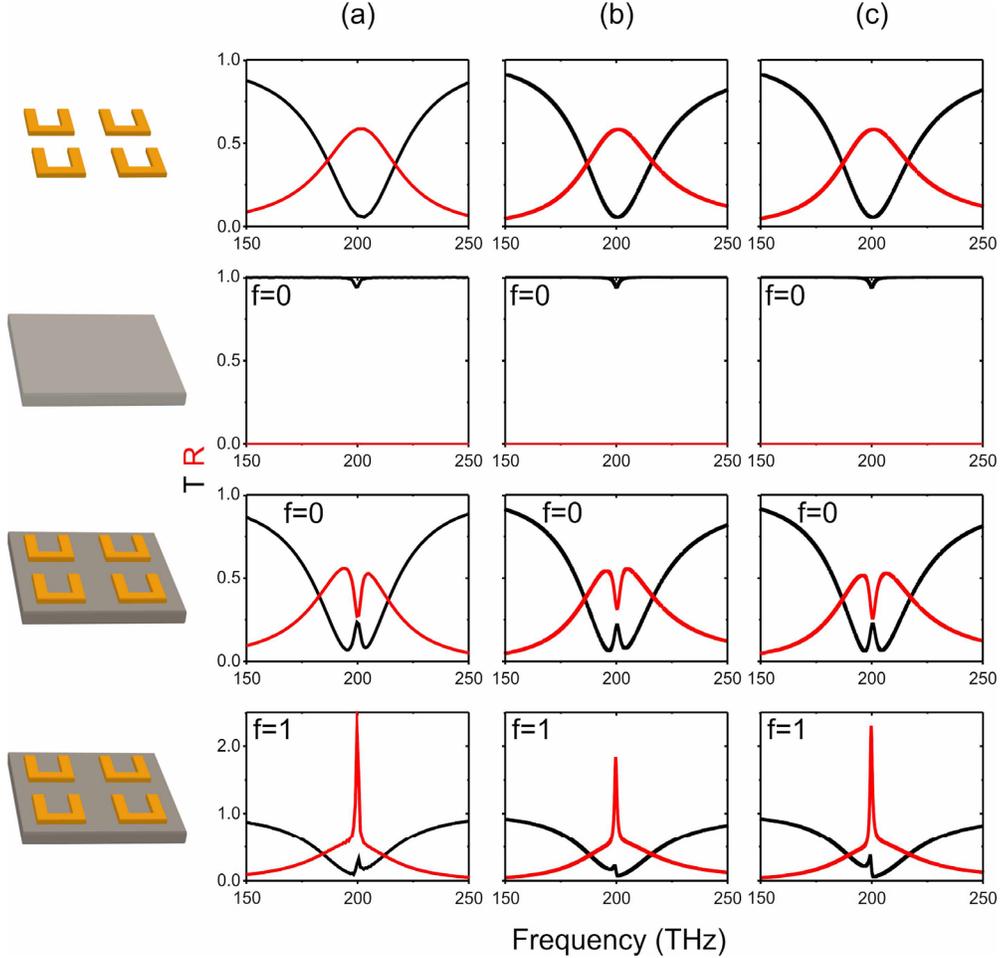

Fig. 2. Normal-incidence intensity transmittance $T$ (black) and reflectance $R$ (red) spectra. From top to bottom row (see schemes on the left-hand side): split-ring resonators only, gain film for $f=0$ only, both combined for $f=0$, and both combined for $f=1$. (a) Complete numerical finite-difference time-domain solutions of the three-dimensional vector Maxwell equations for the geometry depicted in Fig.1. (b) Same for the transfer-matrix treatment of the toy model. (c) Same for the Maxwell-Garnett treatment of the toy model. Model parameters are: $\Omega_{2LS}=2\pi\times200$ THz, $\Omega_{pl}=2\pi\times200$ THz, $\gamma_{2LS}=7.53\times10^{12}$ s$^{-1}$, $\gamma_{pl}=29.5\times10^{12}$ s$^{-1}$, $d_{2LS}=6.5\times10^{-29}$ Cm, $d_{pl}=6.2\times10^{-26}$ Cm, $N_{2LS}=5.05\times10^{23}$ m$^{-3}$, $N_{pl}=5.56\times10^{20}$ m$^{-3}$, $l_{2LS}=50$ nm, $l_{pl}=20$ nm, and $L=3.3416\times10^{10}$ m/F such that $V_{2LS}=7.1\times10^{11}$ s$^{-1}$ and $V_{pl}=6.449\times10^{14}$ s$^{-1}$ result.

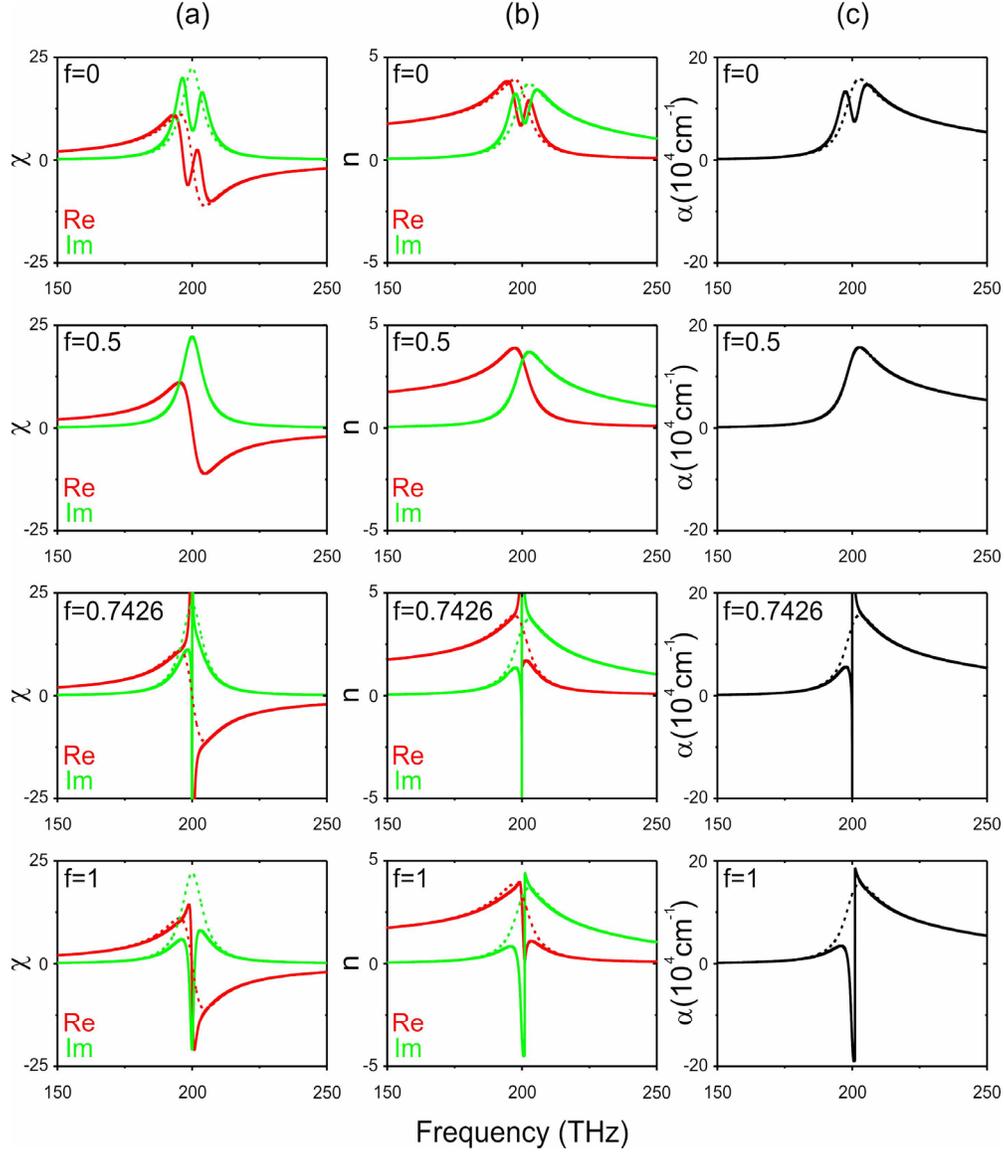

Fig. 3. (a) Toy model complex linear susceptibility $\chi$, (b) complex refractive index $n$, and (c) absorption coefficient $\alpha$ within the Maxwell-Garnett treatment. The solid (dashed) curves are the results with (without) local field coupling $L$. Real (imaginary) parts of complex quantities are red (green). The two-level system occupation $f$ increases from top to bottom row as indicated. Model parameters are identical to those in Fig.2.

Examples for linear optical spectra according to the Maxwell-Garnett approach are given in column (c) of Fig.2. It can be seen that the qualitative agreement with the transfer-matrix calculations shown in column (b) is very good. This is especially true for the lower two rows. The upper two rows of columns (c) and (b) are trivially identical (as they refer to only a single layer each). Furthermore, column (c) in Fig.2 also nicely qualitatively agrees with the complete numerical calculations shown in column (a), although small quantitative deviations do arise. We can conclude that not only the local-field approximation but also the Maxwell-Garnett effective-medium approximation is well justified under these conditions.

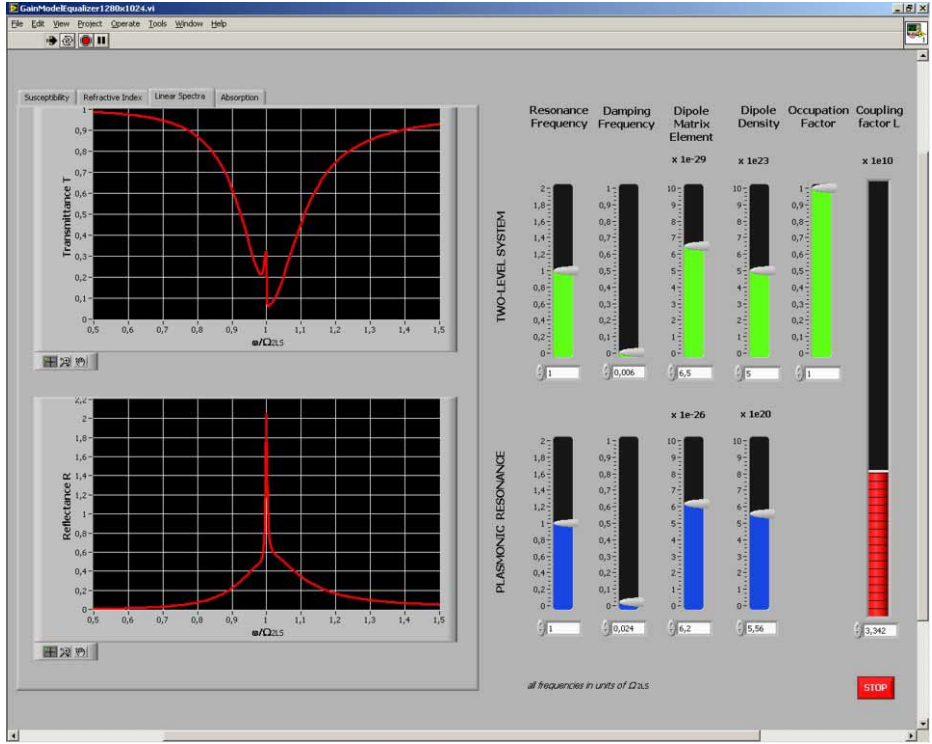

Fig. 4. Readers interested in the numerous parameter combinations other than the few selected special cases shown in Figs.2 and 3 may download a software free of cost [25], allowing to calculate all relevant quantities of our toy model within the Maxwell-Garnett approach. The surface of this software is depicted here. All ten model parameters can be adjusted and the spectra change in real time. Default parameters are those of Figs.2 and 3 and $f$=1. All frequencies and dampings are normalized with respect to the fixed 2LS transition frequency $\Omega_{2LS}=2\pi\times200$ THz (corresponding to about 1.5-µm wavelength).

The corresponding Maxwell-Garnett effective parameters are depicted in Fig.3. Here, the solid curves are the complete results. For the dashed curves, the local-field parameter is artificially set to $L$=0. The comparison of solid and dashed curves, respectively, allows for assessing the influence of the local-field coupling $L$. Generally, the absorption lineshape is Fano-like – an aspect that we have previously pointed out for coupled classical harmonic oscillators [23]. The Fano character becomes especially obvious for $f$=0 and if one of the two resonances exhibits a much larger damping than the other one (not shown here, see Fig.4). More importantly, Fig.3 also shows the evolution of the absorption spectrum for increasing two-level-system upper-state occupation $f$ = 0, 0.50, 0.7426, and 1. As transparency of the 2LS at $f$=0.5 is approached, the avoided crossing obviously collapses. Increasing $f$ further, a sharp feature develops that becomes infinitely sharp if the gain exactly compensates the loss. At this critical point ($f$=0.7426 for the parameters chosen in Figs.2 and 3), we expect lasing (or "spasing" [9] or "lasing spasing" [8]). We will further investigate this critical value for $f$ in the following section. Further increase of $f$ again broadens the spectral features because the gain over-compensates the loss. It is important to note that no gain is observed at all in Fig.3 in case that the local-field coupling is artificially switched off (dashed curves for $L$=0). This behavior is expected on the basis of the intuitive reasoning in section 1 in terms of the *effective* gain rather than the 2LS gain itself that matters. The discontinuities of Im($n$) occur at zeroes of Re($n$), where the sign of the root of the electric permittivity has to change in order to fulfill the condition Re($n$)≥0.

Intuitively, one might have expected that the intensity transmittance $T$ will exceed unity as soon as effective gain occurs (see above). This can be the case, but Figs.2 and 3 show that this is not necessarily the case. At and around the critical point, the real part of the refractive index becomes very large compared to unity or close to zero, either of which leads to a large impedance mismatch with respect to vacuum that gives rise to a large intensity reflectance $R$. As a result, the transmittance $T$ is well below unity and shows no obvious signs of gain at all. We can conclude that transmittance alone is not necessarily a good experimental observable at this point.

As a side remark, we note that this sharp spectral resonance in the real part of the refractive index $n$ for the critical value of the occupation $f$ in Fig.3 implies that the effective group velocity is getting extremely small or even zero in a very narrow spectral range. Notably, one could furthermore envision dynamic tailoring of the effective group velocity via reducing $f$ with respect to the critical value, which is possible by reducing the pump rate $\Gamma_{\text{pump}}$. (Steady-state increase of $f$ with respect to the critical value is not possible, see next section). These small effective group velocities might be interesting in the context of "slow light".

In Fig.2(c) and Fig.3, we have intentionally only shown few selected aspects of the rather rich behavior of the model. In order to allow for a broad overview, we provide the reader with a program free of cost [25] that allows for playing with all parameters of the toy model in real time (see Fig.4). Obviously, each of the two uncoupled resonances has four parameters (center frequency $\Omega$, damping $\gamma$, dipole moment $d$, and density $N$). In addition, the important phenomenological local-field coupling parameter $L$ and the two-level-system occupation $f$ can be freely adjusted – altogether 10 parameters.

## 4. Steady-State Gain and Occupation Pinning

While any value of the 2LS occupation $f \in [0,1]$ can occur under transient conditions, this is not true under steady-state conditions: If the coupled system exhibits a gain that exceeds the loss, stimulated emission will eventually reduce the two-level system upper-state occupation probability $f$ until some steady-state value is reached – a phenomenon, which is well known as gain pinning in the context of a LASER [22,28]. Thus, values of $f$ above that steady-state value must be treated with a grain of salt. Also see Ref.[29] for a corresponding discussion regarding constraints for negative-index metamaterials due to causality.

Mathematically, polarization envelopes which neither grow ("too much gain") nor decay ("not enough gain") obviously need to have constant envelope, i.e., the imaginary part of the corresponding eigenfrequency $\omega$ of the coupled system has to be strictly zero. Without external light field, i.e., for $E=0$ on the RHS of (2) and (3), and for constant pump rate $\Gamma_{\text{pump}}$, the two complex-valued eigenfrequencies of the above coupled equations (2)-(6) are given by

$$\omega = \frac{\Omega_{\text{2LS}} + \Omega_{\text{pl}}}{2} - i\frac{\gamma_{\text{2LS}} + \gamma_{\text{pl}}}{2} \pm \sqrt{\left(\frac{\Omega_{\text{2LS}} - \Omega_{\text{pl}}}{2} - i\frac{\gamma_{\text{2LS}} - \gamma_{\text{pl}}}{2}\right)^2 + V_{\text{2LS}}V_{\text{pl}}(1-2f)}. \quad (11)$$

The behavior of $\omega$ in the complex frequency plane versus $f$ according to (11) is illustrated in Fig.5. Zero imaginary part of (11) immediately translates into the general condition for the 2LS occupation $f$

$$\frac{\gamma_{\text{2LS}} + \gamma_{\text{pl}}}{2} = \pm \text{Im}\left(\sqrt{\left(\frac{\Omega_{\text{2LS}} - \Omega_{\text{pl}}}{2} - i\frac{\gamma_{\text{2LS}} - \gamma_{\text{pl}}}{2}\right)^2 + V_{\text{2LS}}V_{\text{pl}}(1-2f)}\right). \quad (12)$$

If (12) is fulfilled, the denominator of (7) becomes strictly zero for the real frequency $\omega$ resulting from (11), hence the linear optical response (10) diverges at this point.

**Case (a):** Equation (12) has, e.g., a simple transparent special solution for the important degenerate case, i.e., for $\Omega_{2LS}=\Omega_{pl}$. We obtain

$$f = \frac{1}{2}\left(1 + \frac{\gamma_{pl}\gamma_{2LS}}{V_{pl}V_{2LS}}\right) := \frac{1}{2}\left(1 + \frac{\gamma^2}{V^2}\right) \in [0,1]. \tag{13}$$

For example, for the parameters of Fig.2, (13) leads to $f=0.7426$ – which is why we have depicted this particular value in the third row of Fig.3.

Recall that, according to the Pauli exclusion principle, the occupation $f$ in (13) needs to be in the interval [0,1]. Thus, an effective coupling frequency $V$ (= geometric mean of the two couplings) smaller than the effective damping $\gamma$ (= geometric mean of the two dampings) leads to values of $f$ exceeding unity. In other words: We have found a critical threshold value for the strength of the effective local-field interaction $V$. For values below that critical value, no lasing (spasing) action can occur under steady-state conditions. From (11), the (real-valued) lasing SPASER frequency results as $\omega=\Omega_{2LS}=\Omega_{pl}$.

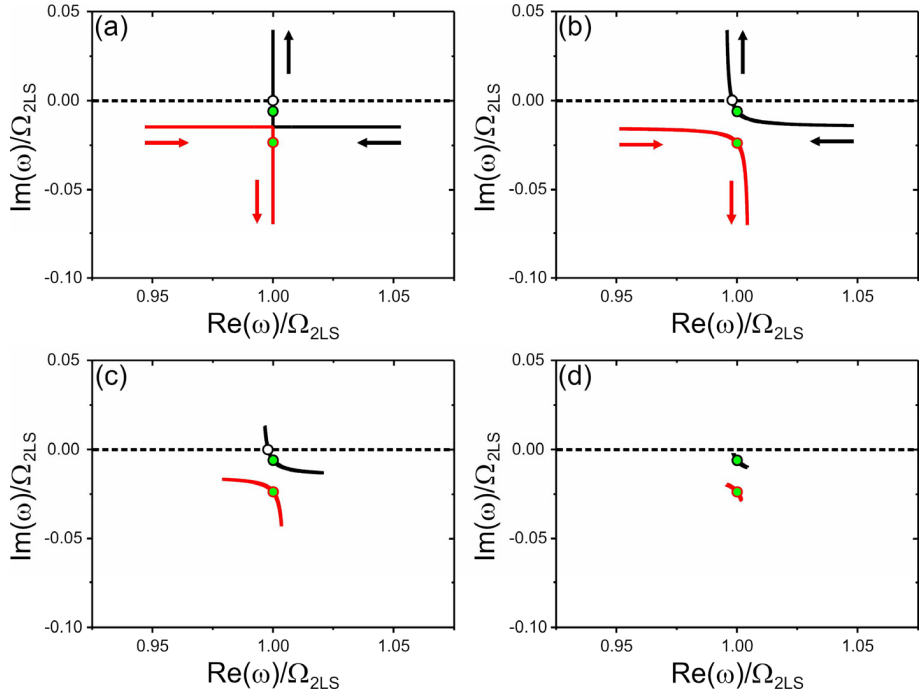

Fig. 5. (a) The two complex eigenfrequencies $\omega$ according to (11). The occupation factor $f$ runs from 0 (no pumping) to 1 (complete inversion) along the direction indicated by the arrows. Model parameters are chosen as in Figs.2 and 3. At the crossing with the real frequency axis, the imaginary part of $\omega$ becomes zero, corresponding to the only possible non-trivial *stationary* solution of the lasing SPASER (see open circle). The occupation $f$ is pinned via this condition (12). The two eigenfrequencies *without* coupling, i.e., for $L=0$, are shown by the green filled circles. Note that these uncoupled complex eigenfrequencies do not depend on $f$ at all. It is instructive to compare the complex eigenfrequencies shown here with the linear optical spectra shown in Figs.2 and 3. (b) Same as (a) but $\Omega_{pl} \rightarrow 0.99 \times \Omega_{pl}$, (c) as (b) but $L \rightarrow L/2$, and (d) as (b) but $L \rightarrow L/5$. For the latter, stationary SPASER action is obviously no longer possible.

The condition of strong coupling $V \geq \gamma$ clearly also implies that an avoided crossing of the two-level-system resonance and the plasmonic resonance can be seen in the linear optical spectra without inversion, i.e., for $f=0$ (see previous section). Hence, within our model, the avoided crossing can be viewed as a prerequisite for obtaining lasing SPASER action (also compare Fig.5).

**Case (b):** Another simple special case of (12) results for identical damping rates, i.e., for $\gamma = \gamma_{2LS} = \gamma_{pl}$, but finite detuning $\Delta\Omega = \Omega_{2LS} - \Omega_{pl} \neq 0$. We derive

$$f = \frac{1}{2}\left(1 + \frac{\gamma^2 + (\Delta\Omega/2)^2}{V^2}\right) \in [0,1]. \tag{14}$$

Clearly, increasing the detuning $\Delta\Omega$ increases the necessary 2LS occupation $f$, until eventually no physical solution $f \in [0,1]$ occurs any more for $\gamma^2 + (\Delta\Omega/2)^2 > V^2$. The detuning simply reduces the optical gain that is accessible for the plasmonic resonance from the two-level system (and so does the damping). From (11), the (real-valued) lasing SPASER frequency results as $\omega = (\Omega_{2LS} + \Omega_{pl})/2$.

Note that – in the entire reasoning of this section – we have deliberately and tacitly neglected any optical feedback due to back-reflections from sample interfaces or even from mirrors due to *propagating* electromagnetic waves (i.e., we have set $E=0$ on the RHS of equations (2) and (3)) because our aim has been to discuss the case of pure *evanescent-field* (or local-field) optical feedback. The same holds true for the following section. Experimentally, this ideal case could, in principle, be realized by appropriate anti-reflection coatings of the structure at the lasing wavelength. It should be clear, however, that there will be a continuous transition from an ideal SPASER to a usual LASER if the relative strength of the far-field optical feedback with respect to the near-field (local-field) optical feedback is increased continuously. Studying this transition is not subject of the present letter.

## 5. Lasing SPASER Relaxation Oscillations

Next, we discuss an example for the time-dependent behavior of our toy model (2)-(6). We consider a pump rate $\Gamma_{pump} = \Gamma_0(1-f)$. The Pauli blocking factor $(1-f)$ acknowledges that the upper state of each two-level system cannot be occupied with more than one electron, i.e., it guarantees $f \in [0,1]$. $\Gamma_0$ shall be zero until time $t=0$. Then it is switched on to a constant value. Also, we seed the polarizations with a tiny but finite value. This is necessary because our toy model does not contain any spontaneous emission whatsoever. As a result, the transition amplitudes would otherwise be strictly zero – no matter how much gain the system develops. Alternatively, one could use a weak external seed pulse of light $E(t)$. None of these details is really important, the initial conditions just need to be non-zero.

Under these conditions, by inserting the local field according to (2) for $E=0$ on the RHS of (2), equation (5) in RWA becomes

$$\dot{f} + \Gamma_{2LS} f = i\left(p_{2LS}^* V_{2LS} p_{pl} - \text{c.c.}\right) + \Gamma_{pump} = -\Gamma_{stim} + \Gamma_{pump}. \tag{15}$$

Here, we have abbreviated the effective rate of stimulated emission $\Gamma_{stim}$. For example, for the degenerate case, the steady-state lasing threshold (where $\Gamma_{stim}=0$) is reached for the pump rate $\Gamma_{pump} = \Gamma_{2LS} f$, with the pinned 2LS occupation $f$ according to (12). As usual, above this threshold pump rate, the effective rate of stimulated emission increases linearly with $\Gamma_{pump}$, i.e., $\Gamma_{stim} = \Gamma_{pump} - \Gamma_{2LS} f$.

Fig.6 shows selected time-dependent numerical solutions of (2)-(6) for parameters corresponding to those in the caption of Fig.2. Trivially, the 2LS occupation $f$ initially grows linearly in time due to the constant generation rate. At some point, the occupation is sufficiently large for obtaining gain (i.e., Im($n$)<0, see section 3). From this point in time on, the effective rate of stimulated emission $\Gamma_{stim}$ increases exponentially. This growth ends when stimulated emission has depleted the gain so much that the system comes back to transparency (i.e., $\Gamma_{stim}\approx 0$ and Im($n$)>0). The resulting sharp emission spikes in Fig.6 have a temporal width on the order of a few picoseconds. After each spike, the population increases again and the gain recovers. Finally, after several oscillations with decreasing amplitude over a time span of hundreds of picoseconds, a steady-state value of $f$ is reached – that does *not* depend on the pump rate $\Gamma_{pump}$. This aspect reflects the *gain pinning* already discussed in the previous section.

The square modulus of the effective polarization (10) shows a behavior closely similar to that of the effective rate of stimulated emission in Fig.6 (hence it is not depicted here).

The damped oscillation scenario is just the counterpart of the relaxation oscillations that are well known from the usual LASER rate equations [22,28]. As usual, the relaxation oscillation frequency increases with increasing pumping level [22,28] – a trend that is also clearly visible in Fig.6. This observation emphasizes that our toy model and the usual LASER rate equations share certain similarities. Section 3, however, has also shown aspects that do not occur in the LASER rate equations at all.

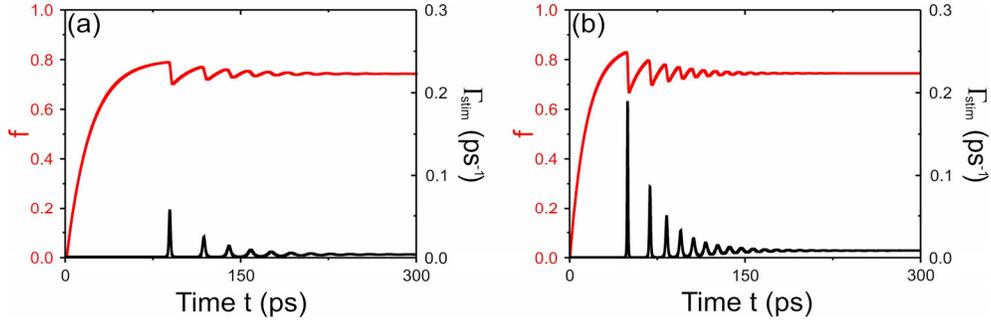

Fig. 6. Switch-on of the lasing SPASER within our toy model leading to pronounced rapid relaxation oscillations of the two-level system occupation $f$ and the effective rate of stimulated emission $\Gamma_{stim}$. The pump rate $\Gamma_{pump}=\Gamma_0(1-f)$ with constant $\Gamma_0$ after time $t=0$ is parameter. The model parameters are identical to those of Figs.2 and 3, $\Gamma_{2LS}=10^{10}$ s$^{-1}$. The lasing SPASER frequency results as $\omega=\Omega_{2LS}=\Omega_{pl}$. (a) $\Gamma_0=4\times 10^{10}$ s$^{-1}$ (just slightly above threshold) and (b) $\Gamma_0=6\times 10^{10}$ s$^{-1}$.

## 6. Conclusions

In conclusion, we have proposed and analyzed a simple and intuitive model for a (bosonic) metamaterial resonance coupled to a (fermionic) two-level-system gain resonance via local-field (evanescent-field) interactions. Especially the presented analytic solutions might help experimentalists in getting a feeling for designing actual SPASER structures. The model contains one purely phenomenological parameter $L$ that describes the strength of the local-field interaction. This parameter needs to be chosen through comparison with numerical calculations of the plasmonic nanostructure via the three-dimensional vector Maxwell equations. The choice of all other model parameters for a given particular configuration is straightforward. To allow the reader for playing with the altogether ten model parameters, we provide a corresponding software [25]. Next, analytic results for steady-state gain/occupation pinning have been derived. Furthermore, we have presented numerical solutions for the time-dependent problem that exhibits the usual laser relaxation oscillations. The latter aspect shows that our toy model for the SPASER shares similarities with the well established LASER rate equations.

Clearly, our simple modeling leaves plenty of room for future improvements. For example, proper treatment of the semiconductor will require accounting for Coulomb interaction effects via the semiconductor Bloch equations [22,28]. Furthermore, self-consistent solutions of the material and the Maxwell equations should include a spatially inhomogeneous response of the gain material in the vicinity of the plasmonic nanostructure.

Broadly speaking, we have seen that one must not assume that gain can be added to a metamaterial just to reduce the losses and leave the metamaterial properties (e.g., magnetic permeability or negative refractive index) unaltered otherwise. Strong coupling to a gain resonance inherently and unavoidably changes the system, resulting in a new effective system with new effective properties that need to be evaluated.

**Acknowledgements**

We thank Hyatt Gibbs, Galina Khitrova, and Wolfgang Stolz for discussions. We acknowledge financial support provided by the Deutsche Forschungsgemeinschaft (DFG) and the State of Baden-Württemberg through the DFG-Center for Functional Nanostructures (CFN) within subproject A1.5. The project PHOME acknowledges the financial support of the Future and Emerging Technologies (FET) programme within the Seventh Framework Programme for Research of the European Commission, under FET-Open grant number 213390. Also, we acknowledge funding through the METAMAT project by the Bundesministerium für Bildung und Forschung (BMBF). The research of S.L. is further supported through a "Helmholtz-Hochschul-Nachwuchsgruppe" (VH-NG-232), the PhD education of N.M. and M.R. through the Karlsruhe School of Optics & Photonics (KSOP). J.L.G.-P. acknowledges support by the I3P-CSIC grant program.